\documentclass[utf8]{frontiersSCNS} 

\usepackage{url,hyperref,lineno,microtype,subcaption}

\def\keyFont{\fontsize{8}{11}\helveticabold }
\def\firstAuthorLast{Rice and Yeates} 
\def\Authors{Oliver E. K. Rice and Anthony R. Yeates\,$^{*}$}
\def\Address{Department of Mathematical Sciences, Durham University, Durham, DH1 3LE, UK}
\def\corrAuthor{Corresponding Author}

\def\corrEmail{anthony.yeates@durham.ac.uk}

\begin{document}
\onecolumn
\firstpage{1}

\title[Eruptivity Criteria for Two-dimensional Magnetic Flux Ropes]{Eruptivity Criteria for Two-dimensional Magnetic Flux Ropes in the Solar Corona} 

\author[\firstAuthorLast ]{\Authors} 
\address{} 
\correspondance{} 

\extraAuth{}

\maketitle

\begin{abstract}

\section{}
We apply the magneto-frictional approach to investigate which quantity or quantities can best predict the loss of equilibrium of a translationally-invariant magnetic flux rope. The flux rope is produced self-consistently by flux cancellation combined with gradual footpoint shearing of a coronal arcade which is open at the outer boundary. This models the magnetic field in decaying active regions on the Sun. Such a model permits two types of eruption: episodic small events caused by shearing and relaxation of the overlying arcade, and major eruptions of the main low-lying coronal flux rope. Through a parameter study, we find that the major eruptions are best predicted not by individual quantities but by thresholds in the ratios of squared rope current to either magnetic energy or relative magnetic helicity. We show how to appropriately define the latter quantity for translationally-invariant magnetic fields, along with a related eruptivity index that has recently been introduced for three-dimensional magnetic fields. In contrast to previous configurations studied, we find that the eruptivity index has only a weak predictive skill, and in fact is lower prior to eruption, rather than higher. This is because the overlying background magnetic field has the same direction as the arcade itself. Thus we propose that there are a whole class of solar eruptions that cannot be predicted by a high eruptivity index.

\tiny
 \keyFont{ \section{Keywords:} Sun, Solar corona, Magnetic Fields, Eruptions, Helicity} 
\end{abstract}

\section{Introduction}

Flux ropes are twisted bundles of magnetic flux in the solar corona \citep{2020RAA....20..165L}. Accurately predicting their behaviour is essential for reliable space weather predictions, as unstable flux ropes can erupt and lead to large coronal mass ejections \citep[CMEs;][]{2006SSRv..123..251F, 2011LRSP....8....1C}. The causes of such eruptions are not yet indisputably understood and a variety of mechanisms have been proposed. These have been explored with approaches ranging from analytic two-dimensional models to three-dimensional full magnetohydrodynamic (MHD) simulations.

Analytic two-dimensional models of flux rope behaviour date back to \citet{1974A&A....31..189K} and  \citet{1978SoPh...59..115V}, who modelled a horizontal line current and its interaction with a specified background magnetic field. They established conditions on the current that allow it to be stable, and showed that, for an eruption to occur, the background magnetic field strength must decrease rapidly with increasing height.

Further analytic approaches have introduced the torus instability \citep[e.g.,][]{2006PhRvL..96y5002K} where the flux rope is modelled as a current ring rather than a line current. Rather than a condition on the decay of the background field with altitude, it is instead proposed that for instability to occur the background field component orthogonal to the torus must decrease sufficiently quickly. It has since been shown that the conditions for such an instability are essentially the same as for instability of a line current, and they are just two special cases of a continuous theory of more general current paths \citep{2014ApJ...789...46K}. Although these works provide useful theoretical background as to the nature of flux rope behaviour, they are difficult to relate to physical ropes as these tend to be significantly more complex and much of the physics is by necessity not taken into account. There are indeed other conflicting explanations for instability, such as that of \citet{2017ApJ...843..101I}, who propose that fast magnetic reconnection can occur below a flux rope even if the overlying magnetic field does not decay with altitude.

Full 3D MHD simulations of flux rope formation and behaviour are also now possible. \citet{2013ApJ...778...99L, 2014ApJ...787...46L} presented a series of such simulations, based on magnetic flux emergence into a preexisting field. Some ropes were found to be eruptive and others stable, depending on the configuration of the background magnetic field. \citet{2017A&A...601A.125P} then analysed various properties of the system for these simulations, looking for properties that could predict whether or not the flux rope would erupt. Most diagnostics, such as the magnetic free energy and relative magnetic helicity, were shown not to have any strong correlation to eruptivity. However, they identified an ``eruptivity index'' -- the fraction of the relative magnetic helicity that is in the current-carrying component -- which became large only in erupting cases. In a different set of numerical simulations, where coronal flux ropes are created by  a variety of solar surface footpoint motions, this eruptivity index again increased prior to eruption and the eruption was found to occur for a fixed value of the index, irrespective of the pattern of footpoint motions that injected the energy \citep{2018ApJ...863...41Z}. However, while promising, a complete theoretical understanding of this eruptivity index and its generality remains lacking.

Our modeling approach is chosen as a compromise between the simple analytic models of the past and the expensive full MHD simulations now possible. We use the magneto-frictional model, pioneered by \citet{1986ApJ...309..383Y}, whereby a fictional velocity field is determined explicitly from the magnetic field as opposed to using the fluid equations.
For tracking the quasi-static injection of magnetic energy into the corona through solar surface motions, the model provides a viable alternative to full MHD simulations, at a fraction of the computational cost.
Magneto-frictional models have previously been coupled with time-dependent lower boundary conditions based on imposed surface flux transport \citep[e.g.,][]{2008SoPh..247..103Y} and realistic flux rope formation, as well as loss of equilibrium, is observed in such simulations \citep{2006ApJ...641..577M, 2009ApJ...699.1024Y, 2017ApJ...846..106L, 2020ApJS..250...28H}. Flux ropes are also formed on a smaller scale when the model is driven by high-resolution observations within active regions \citep[e.g.,][]{2018ApJ...852...82Y}.
Since the magneto-frictional model neglects the full equation of motion, it cannot accurately follow the dynamics once a flux rope loses equilibrium and erupts. Nevertheless, if ropes produced by magneto-friction are used to initialise full MHD simulations, it is found that the magnetic flux ropes do indeed lose equilibrium at the same point, and can lead to CME eruptions \citep{2013ApJ...779..129K, 2013A&A...554A..77P}. Thus magneto-friction can act as an accurate model for the pre-eruption evolution.

In our study, we use the magneto-frictional approach but simplify it to a 2.5-dimensional (translationally invariant) cartesian domain, which significantly increases computational speed while still exhibiting the fundamental features of fully 3D models. In particular, both the flux rope and the (sheared) overlying arcade are formed self-consistently by flux cancellation and shearing motions on the photosphere. In our model these shearing motions are purely large-scale, essentially modelling the differential rotation. As such, we focus on the more gradual evolution in decaying active regions -- a potentially different scenario to the active region eruptions where the eruptivity index was previously studied, but nevertheless an important source of solar eruptions.

In a similar manner to the work of \citet{2017A&A...601A.125P}, we attempt to find a scalar quantity that acts as a proxy for the eruptivity of our 2.5D flux rope. The large number simulations we are able to run (in the order of $500$) allows for an extensive parameter study based on the variation of the magneto-friction coefficient and the rate of photospheric flux cancellation. Using a probabilistic approach, a large number of eruptive and non-eruptive ropes will be compared against diagnostic measurements of the system at various points of their evolution. We note that, in addition to the formation of flux ropes, our model exhibits ``arcade eruptions'' whereby we observe periodic reconnection at the top of a sheared magnetic arcade. It is possible that these represent streamer blowouts or even ``stealth CMEs'', as coined by \citet{2012LRSP....9....3W}, since they are characterised by the lack of a detectable signature on the solar surface. The nature and period of these eruptions (around every 25-30 days) matches well between our 2.5D model, global magneto-frictional simulations and observations \citep{bhowmik2021}. However, the main focus of this paper is not on arcade eruptions but ``flux rope eruptions'', when we observe magnetic reconnection below a flux rope and the rope itself moves upwards out of the domain. Such eruptions are larger than arcade eruptions and it is likely that in reality a significant number of CMEs occur as a result of such a mechanism.

We begin in Section \ref{sec:approach} by outlining the mathematical basis of our model, including the mechanism by which we represent surface shearing (influenced by the differential rotation of the sun at different latitudes), photospheric diffusion and the effect of the solar wind. We identify the variable parameters in the model and discuss which of these we can use to produce an array of differing flux rope behaviour. We then discuss the system diagnostics we have chosen to focus on, including a newly-defined measure of the relative helicity in two dimensions (Section \ref{sec:hr}). This measure allows us to compare our results against 3D equivalents and calculate the eruptivity index. The results of our parameter study are presented in Section \ref{sec:results}, and the findings discussed in Section \ref{sec:discussion}.

\section{Methods} \label{sec:approach}

\subsection{Magneto-frictional model}

Rather than using full magnetohydrodynamics (MHD), we adopt a simplified approach that has been developed for global modelling of the solar coronal magnetic field: the magneto-frictional model \citep[e.g.,][]{2012LRSP....9....6M}. On a global scale, this technique has been applied with time-dependent lower boundary conditions from surface flux transport models \citep[e.g.,][]{2008SoPh..247..103Y,2014SoPh..289..631Y}. We adopt a 2.5-dimensional (translationally invariant) version of this approach.

In full MHD models, the velocity field, ${\bf v}$, is determined by the momentum equation
\begin{equation}
\rho\frac{\mathrm{D}{\bf v}}{\mathrm{D}t} = {\bf j}\times{\bf B} - \nabla p - \rho\nabla\Psi,
\end{equation}
where $\rho$ is the fluid density, $p$ is the fluid pressure, $\Psi$ is the gravitational potential, and ${\bf j} = \mu_0^{-1}\nabla\times{\bf B}$ is the current density. This is coupled to the induction equation
\begin{equation}
\frac{\partial{\bf B}}{\partial t} = \nabla\times({\bf v}\times{\bf B}) - \eta \nabla \times \bf j,
\label{eqn:ind}
\end{equation}
along with additional fluid equations to close the system. In the magneto-frictional method, equation \eqref{eqn:ind} is retained, but the inertial terms, pressure gradients and gravity are neglected and instead a frictional velocity is imposed as
\begin{equation}
\nu{\bf v}= (\nabla\times{\bf B})\times{\bf B}.
\label{eqn:vxb}
\end{equation}
Combined with the induction equation, this leads to monotonic relaxation towards a stationary force-free field with ${\bf j}\times{\bf B}={\bf 0}$. The friction coefficient $\nu$ is typically given the form $\nu=\nu_0|{\bf B}|^2$ (with some minimum value imposed) so that the overall evolution is independent of the magnitude of ${\bf B}$ and relaxation is not unduly slow near to magnetic null points \citep{1986ApJ...309..383Y}.

In the outer corona, the solar wind outflow prevents the magnetic field from being force-free, but this effect can be approximated in the magneto-frictional model by relaxing towards an equilibrium with a specified outflow ${\bf v}_{\rm out}$, thus choosing ${\bf v}$ according to
\begin{equation}
\bf v = \frac{(\nabla\times{\bf B})\times{\bf B})}{\nu} + {\bf v}_{\rm out} \label{eqn:vxbout}
\end{equation} 
This \textit{ad hoc} approach was introduced by \citet{2006ApJ...641..577M}, and has subsequently been used in global magneto-frictional models of solar and stellar coronae \citep[e.g.][]{2014SoPh..289..631Y, 2016MNRAS.456.3624G, 2018ApJ...869...62M, 2020SoPh..295..101M}. 

\subsection{Our 2.5-Dimensional Model}

In our model, we simplify the problem by using a Cartesian coordinate system, and removing any dependence of the solution on the $y$ coordinate. The domain ratio is taken to be $(x:z) = (2:1)$. The resulting 2.5-dimensional system captures many of the essential features of the evolution in the lower solar corona, while affording a vast reduction in computational expense. This has allowed us to run an extensive parameter study, comprising hundreds of simulations. 

The state of the system can be described in terms of a vector potential  ${\bf A}(x,z,t)$, such that 
\begin{equation}
    {\bf B}(x,z,t) = \nabla \times {\bf A}(x,z,t).
\end{equation}
The vector field ${\bf A}$ is then evolved according to
\begin{equation}
\frac{\partial{\bf A}}{\partial t} = -{\bf E},
\label{eqn:A}
\end{equation}
where ${\bf E}$ is the electric field, satisfying Ohm's Law
\begin{equation}
    {\bf E} = \eta {\bf j} - {\bf v} \times {\bf B},
\end{equation}
in terms of the magnetic field ${\bf B}$, current density ${\bf j}$ and frictional velocity ${\bf v}$ as given in equation \eqref{eqn:vxbout}. Here $\eta$ is a constant representing coronal turbulent diffusivity, which is assumed to be much more significant than ohmic diffusivity in the highly-conducting corona, though still smaller than the other effects in the model.
The outflow velocity used to represent the solar wind is taken to be
\begin{equation}
    {\bf v}_{\rm out} = v_1\left(\frac{z}{z_1}\right)^{10}{\bf{e}}_z,
\end{equation}
so the effect is minimal near the lower boundary and increases rapidly near the top boundary at $z=z_1$ at which the parameter $v_1$ gives the maximum speed.

\begin{figure}[ht]
\includegraphics[scale=0.97]{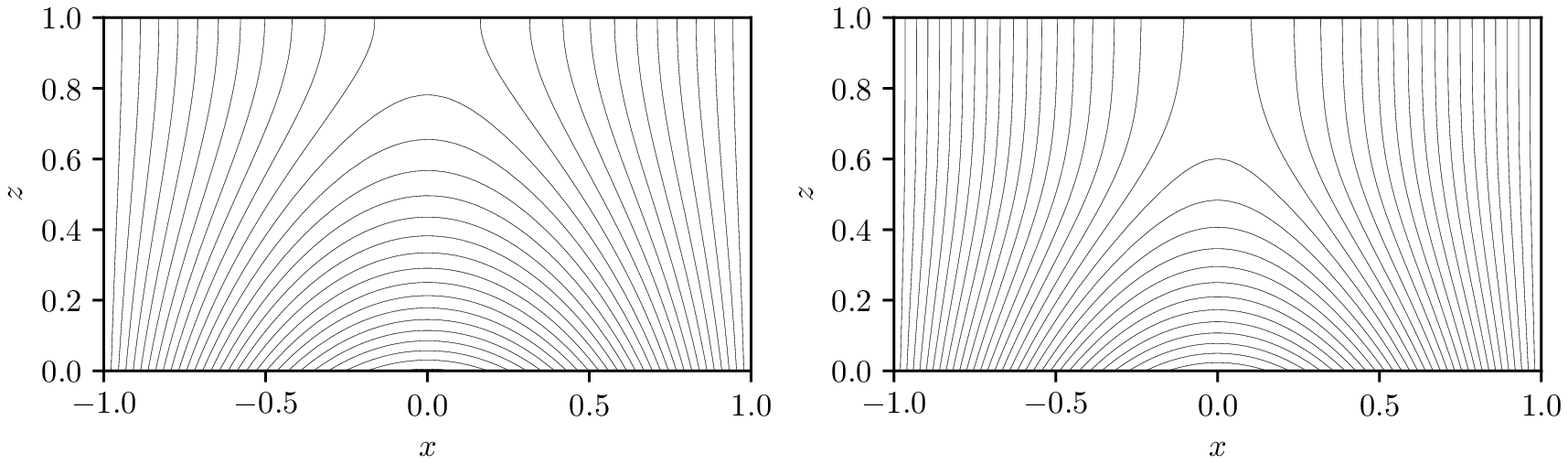}
\caption{Examples of initial conditions for the 2.5D magneto-frictional code in a cartesian domain. The left pane is a potential field and the right pane is an equivalent outflow field, with the same lower boundary condition. The black lines represent magnetic field lines.}
\label{fig:inits}
\end{figure}

The simulations are initialised with an ``outflow field'' \citep{2021arXiv211001319R}, a variation on a potential arcade that takes into account the effect of ${\bf v}_{\rm out}$, such that the system is initially in equilibrium. As seen in Figure \ref{fig:inits} these fields are similar to potential arcades, except in the upper half of the domain where the field lines open out to become more vertical.

The boundary conditions determine the overall behaviour of the system. On the top boundary ($z=z_1$) we have the condition that ${\bf B} \perp {\bf n} = 0$, ensuring that the magnetic field lines are vertical/radial here. This condition is consistent with observations that the field lines in the real corona become almost radial above a certain altitude. The notable exception to this is during eruptions, but these last a relatively short time and occur in our model even with the radial condition imposed. On the sides of the domain we set ${\bf B} \cdot {\bf n} = 0$, i.e. there is no magnetic flux through the sides. The lower boundary condition is more complex, and incorporates two effects: photospheric shearing and photospheric diffusion.

In our model, the photospheric shearing is assumed to originate from the differential rotation of the solar surface, namely that the Sun rotates more quickly at the equator than the poles. A magnetic arcade that has footpoints at different latitudes will thus be sheared, stretching the field lines along the polarity inversion line (PIL). We represent this shearing by imposing a velocity in the out-of-plane ($y$) direction at the lower boundary, following the profile 
\begin{equation}
{\bf v} (x,0) = (0, V_{y0}, 0), \qquad V_{y0} = \sin(\pi x).
\label{eqn:v0}
\end{equation}
This profile is chosen as it ensures symmetry and simplicity while approximating the qualitative effect of differential rotation.
The photospheric diffusion, $\eta_0$ represents the large-scale effect of supergranular flows, and is also imposed on the lower boundary \citep{2006ApJ...641..577M}. 
The combined effect of both shearing and diffusion results in the lower boundary condition
\begin{equation}
{\bf E}(x,0) = - V_{y0}(x,0) B_z(x,0){\bf e}_x  -\frac{\eta_0}{\mu_0}\frac{\partial B_z(x,0)}{\partial x}{\bf e}_y .
\end{equation} 
The effect of this is to bring the arcade footpoints closer together, eventually reconnecting to form a twisted flux rope \citep{1989ApJ...343..971V}. Thus the ropes form self-consistently, without the need for any imposed flux emergence. 

Numerically, equation \eqref{eqn:A} is solved on a finite-difference staggered grid \citep{1966ITAP...14..302Y} similarly to \citep{2012ApJ...757..147C}, with typical resolutions of $256 \times 128$ cells. The code is initialised with Python but runs using Fortran 90, making use of MPI parallelisation.

\subsection{Parameters and Units}

There are a number of free parameters in our model that can be easily varied. The aim is to produce a set of simulations that model realistic behaviour well, but provide enough variation to be able to make predictions of future behaviour. The parameters in the model are:
\begin{itemize}
    \item $\eta$ : The coronal diffusion.
    \item $\nu_0$ : The magneto-frictional friction coefficient.
    \item $v_1$ : The imposed solar wind outflow speed.
    \item $\eta_0$: The photospheric diffusion.
    \item $V_{y0}$: The photospheric shearing velocity.
\end{itemize}
However, in our analysis we fix some of these parameters. Firstly, we  take the coronal diffusion to be very small, $\eta \approx 1 \times 10^{-6}$, so that its effect on short-term flux rope behaviour is negligible. (The corresponding diffusion time across the height of the domain would be $10^6$ time units.) Secondly, we effectively set the time unit by fixing the maximum shearing velocity $V_{y0}$ to unity. Finally, we fix the outflow speed $v_1=50$. This is reasonable since the outflow velocity has little effect on flux rope behaviour; the solar wind mainly serves to induce currents in the upper corona, at the top of the domain and far from the flux rope formation region. The only notable effect of an increased outflow velocity is a slight increase in the frequency of arcade eruptions. We choose $v_1 = 50$ to reflect the fact that the solar wind outflow is faster than the shearing velocity from differential rotation. 

This leaves two remaining parameters: $\nu_0$ and $\eta_0$. These can be varied considerably (between certain bounds) and still exhibit realistic behaviour. By running simulations with different combinations of these parameters we observe different flux rope behaviours, with a good mix of eruptive and non-eruptive simulations. The results from this parameter study are described in Section \ref{sec:results}. 

Note that we adopt dimensionless units throughout this paper, with a maximum photospheric shearing velocity of unity according to equation \eqref{eqn:v0}. For comparison to the real corona, one would choose the length unit -- equivalently the height of the domain, $z_1$ -- and specify an observed shearing velocity caused by the differential rotation of the coronal arcade. These would then fix the time that corresponds to one dimensionless time unit in our paper. For example, if we take the angular velocity of differential rotation at latitude $\lambda$ on the Sun to be $\Omega(\lambda) = 0.18 - 2.396\sin^2\lambda - 1.787\sin^4\lambda$ degrees per day \citep{1983ApJ...270..288S} and choose the latitudinal limits of our domain to be $10^\circ$ to $40^\circ$, this results in a maximum shearing velocity $|v_\phi| \approx 0.086\,\mathrm{km}\,\mathrm{s}^{-1}$. Taking $z_1 = 1.8 \times 10^{5}\, \mathrm{km}$ (half of the the latitudinal extent) would then imply that a code time unit is of the order $\approx 25 \, \mathrm{days}$.

\subsection{System Diagnostics}
\label{sec:diags}

In this section we briefly describe the diagnostic measurements of the system used to identify events and ultimately try to make predictions of future behaviour such as eruptions. There are innumerable measurements that could  theoretically be taken of the state of the magnetic field, but for practical purposes we have selected nine, which are as follows:

\subsubsection{Open Flux}

The open flux is the sum of the unsigned magnetic flux through the top boundary of the domain. (Due to the symmetry of the system, the sum of the signed flux is zero.) The open flux increases due to the effect of the solar wind, as the field lines become more vertical near the top of the domain and fewer of them loop back down in the arcade.

\subsubsection{Maximum Current}

For practical computation we set $\mu_0=1$ and define the current density as the curl of the magnetic field: ${\bf j} = \nabla \times {\bf B}$. As a diagnostic, we take the maximum value that this attains in the domain. A potential field has zero current and an outflow field only has current concentrated near the top boundary, due to the effect of the solar wind. 

\subsubsection{Rope Current}

We also measure the integral of the current within the rope, in the direction of the rope axis (the $y$ direction). In our two-dimensional model the rope is easily identifiable as it consists of the infinitely-long field lines that never reach either the photospheric or outer boundary. As the rope is in general in the lower half of the domain, this current is unaffected by the behaviour at the top boundary.

\subsubsection{Magnetic Energy}

The magnetic energy is defined in our unitless system as $E_{\rm M} = \int_0^1\int_{-1}^1\frac{1}{2}|{\bf B}|^2\,\mathrm{d}x\mathrm{d}z$. A potential arcade corresponds to a minimum energy solution for given normal-field boundary conditions, and an outflow field has only a slightly increased energy (less than 1\%). There is a large increase in magnetic energy during flux rope formation as the system evolves far from potentiality. 

\subsubsection{Free Magnetic Energy}

In addition to the overall magnetic energy, we can calculate the ``free magnetic energy'', defined as f $E_{\rm F} = E_{\rm M} - \int_0^1\int_{-1}^1\frac{1}{2}|{\bf B}_P|^2\,\mathrm{d}x\mathrm{d}z$, where ${\bf B}_P$ is a potential magnetic field as calculated in Section \ref{sec:relhel} below. This quantity has the advantage that the contribution to the overall energy from the background magnetic field is lessened, such that the contribution from the non-potential flux rope is more significant.

\subsubsection{Poloidal Rope Flux}

This is a measure of the poloidal (in-plane) magnetic flux contained within the flux rope (the region with infinitely-long field lines), defined as the flux intersecting a chord between the centre of the rope and the edge of the rope (usually the lower edge of the domain).

\subsubsection{Axial Rope Flux}

This is defined as the integral of the magnetic flux in the rope (the region with infinitely-long field lines) in the $y$ direction, parallel to the axis of the rope itself. This appears to be roughly correlated to the rope current.

\subsubsection{Relative Helicity} \label{sec:hr}
\label{sec:relhel}
The classical helicity within a volume $V$ would be defined as $h(V) = \int_V {\bf A} \cdot {\bf B} \, dV, $ where ${\bf A}$ is the vector potential of ${\bf B}$. This quantity is dependent in general on the gauge of ${\bf A}$, and so we use the alternative relative helicity instead \citep{1984JFM...147..133B}. In a 3D domain, this would be calculated by finding a potential field ${\bf B}_P$ matching the original magnetic field on the boundary, and a corresponding vector potential, ${\bf A}_P$. The relative helicity would then be 
\begin{equation}
    H_R = \int_V ({\bf A + A}_P) \cdot ({\bf B - B}_P) \, dV.
    \label{eqn:hr3d}
\end{equation}

Care is required to define the relative helicity for our 2.5D field. A two-dimensional helicity measure for $h(V)$ has been proposed before \citep{1997SoPh..170..283H}, but we are not aware of a previously published two-dimensional analogue for the relative helicity.

We start by considering the 3D formula \eqref{eqn:hr3d} on a finite volume $V_{y_1}$, where $-1<x<1$, $-y_1<y<y_1$ and $0<z<1$. Although ${\bf B}(x,z)$ in our 2.5D field is independent of $y$, the corresponding potential reference field ${\bf B}_P$ will, in general, vary in the $y$ direction. This arises from the fact that it is potential, coupled with the need to match $B_{Py}(x,\pm y_1, z) = B_y(x,z)$ on $y=\pm y_1$.  We define the relative helicity per unit length to be
\begin{equation}
H_R^{(2.5D)} = \lim_{y_1\to\infty}\frac{1}{2y_1}\int_{V_{y_1}}({\bf A} + {\bf A}_P)\cdot({\bf B} - {\bf B}_P)\,\mathrm{d}V, \label{eqn:hr2d}
\end{equation}
where ${\bf B}_P$ and its vector potential ${\bf A}_P$ are calculated on $V_{y_1}$. However, a physically meaningful helicity measure for our 2.5D field cannot possibly require integration in $y$. We will show that $H_R^{(2.5D)}$ not only converges as $y_1\to\infty$ but can indeed be calculated by a two-dimensional integral in $x$ and $z$.

To do this, we decompose ${\bf B}_P$ into three components,
\begin{equation}
{\bf B}_P=\nabla\phi_0(y) + \nabla\phi_1(x,z) + \nabla\phi_2(x,y,z),
\end{equation}
where the first component is a uniform field accounting for the net flux in the $y$ direction,
\begin{equation}
\phi_0 = \Phi_0y, \qquad \Phi_0 = \frac{1}{2}\int_0^1\int_{-1}^1 B_y(x,z)\,\mathrm{d}x\mathrm{d}z,\\
\end{equation}
and the other two components are both potential fields satisfying $\Delta\phi_1=\Delta\phi_2=0$, with corresponding boundary conditions
\begin{align}
&\frac{\partial\phi_1}{\partial x}(\pm1,z) = 0, \qquad \frac{\partial\phi_1}{\partial z}(x,0) = B_z(x,0), \qquad \frac{\partial\phi_1}{\partial z}(x,1) = B_z(x,1),\\
&\frac{\partial\phi_2}{\partial x}(\pm1,y,z) = \frac{\partial\phi_2}{\partial z}(x,y,0)=\frac{\partial\phi_2}{\partial z}(x,y,1)=0, \qquad \frac{\partial\phi_2}{\partial y}(x,\pm y_1,z) = B_y(x,z) - \Phi_0.
\end{align}
Notice that $\nabla\phi_0$ and $\nabla\phi_1$ are independent of both $y$ and $y_1$. The $y$ dependence is concentrated only in $\nabla\phi_2$.

\begin{figure}[ht]
\includegraphics[scale=0.97]{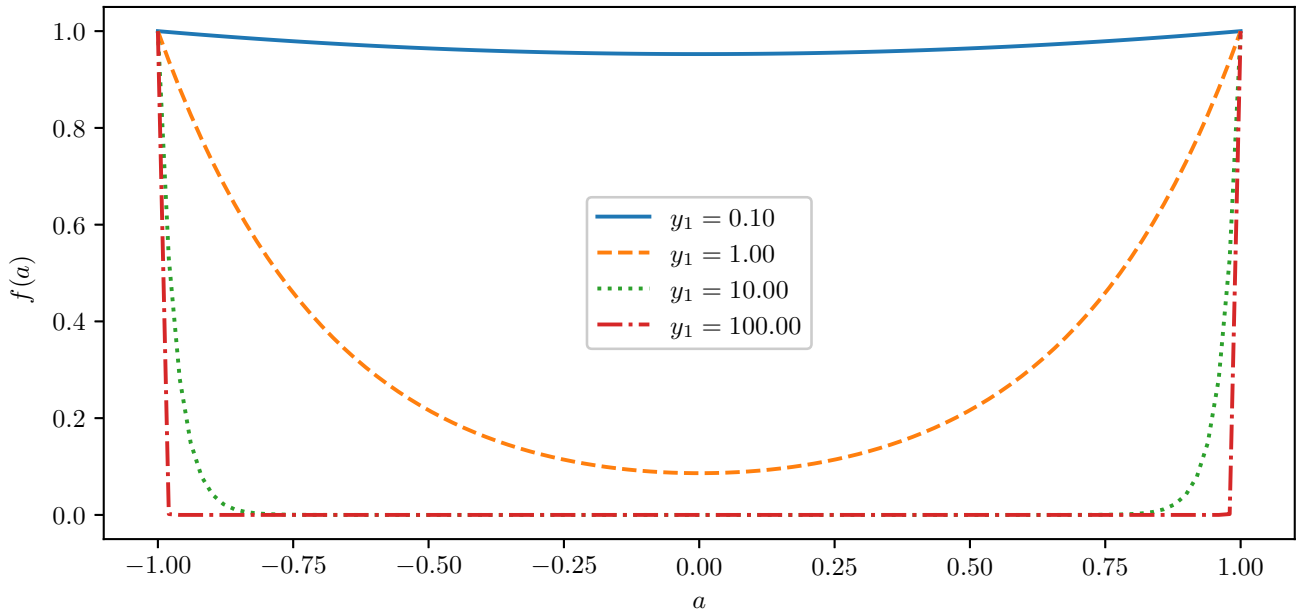}
\caption{Plot of the function $f(a) = \frac{\cosh(k \pi ay_1)}{\cosh(k \pi y_1)}$ for $k=1$ and various values of $y_1$.}
\label{fig:coshplot}
\end{figure}

The potential $\phi_2$ has the important property that, as $y_1$ increases, it becomes more and more concentrated near to the end boundaries $y=\pm y_1$, irrespective of $B_y(x,z)$.
To see this, note that in the Cartesian domain $V_{y_1}$, the solution for $\phi_2$ may be written as a Fourier series
\begin{equation}
\phi_2(x,y,z) = \sum_{m,n\neq 0}c_{m,n}\cos\left(\frac{m\pi(x+1)}{2}\right)\cos\left(n\pi z\right)\sinh(k\pi y),  \label{eqn:phi2}
\end{equation}
where $m$, $n$ are integers, $k^2 = (m/2)^2 + n^2$, and the sum includes all terms except $m=n=0$ (which has been separated as the $\phi_0$ component). The coefficients are then determined by the boundary condition on $\partial\phi_2/\partial y$, which gives
\begin{equation}
c_{m,n} = \frac{2}{k\pi\cosh(k\pi y_1)}\int_0^1\int_{-1}^1B_y(x,z)\cos\left(\frac{m\pi(x+1)}{2}\right)\cos\left(n\pi z\right)\,\mathrm{d}x\mathrm{d}z. \label{eqn:cmn}
\end{equation}
Now consider the value of $\phi_2(x,ay_1,z)$ for some fixed fraction $|a|<1$. Then
\begin{equation}
\phi_2(x,ay_1,z) = \sum_{m,n} \frac{2\sinh(k\pi ay_1)}{k\pi
\cosh(k\pi y_1)}F(x,z),
\end{equation}
where $F(x,z)$ contains the $x$ and $z$ dependence from \eqref{eqn:phi2} and \eqref{eqn:cmn}. Noting that 
\begin{equation}
\lim_{y_1\to\infty}\frac{\sinh(k\pi ay_1)}{\cosh(k\pi y_1)} = \lim_{y_1\to\infty}\frac{\cosh(k\pi ay_1)}{\cosh(k\pi y_1)} = 0,
\end{equation}
we see that the non-zero part of $\nabla\phi_2$ becomes an increasingly smaller fraction of the domain length as $y_1\to\infty$. This is illustrated in Figure \ref{fig:coshplot}. It follows that the contribution from $\nabla\phi_2$ to $H_R^{(2.5D)}$ in equation \eqref{eqn:hr2d} vanishes in the limit, so that 
\begin{equation}
H_R^{(2.5D)} = \int_0^1\int_{-1}^1({\bf A}+{\bf A}_P^{(2.5D)})\cdot({\bf B} - {\bf B}_P^{(2.5D)})\,\mathrm{d}x\mathrm{d}z,
\label{eqn:hr2dxz}
\end{equation}
where ${\bf B}_P^{(2.5D)}(x,z) = \nabla\phi_1(x,z) + \Phi_0{\bf e}_y$. The gauge of ${\bf A}_P^{(2.5D)}$ does not affect the integral (as usual for $H_R$), so one is free to choose an  ${\bf A}_P^{(2.5D)}$  that is independent of $y$ and hence evaluate $H_R^{(2.5D)}$ with a purely 2D integral. This is our approach in this paper.

\subsubsection{Eruptivity Index}
The 3D expression that has been proposed \citep{2017A&A...601A.125P} as an eruptivity index is $\lvert H_J/H_R \rvert$, where
\begin{equation}
    H_J = \int_V ({\bf A - A}_P) \cdot ({\bf B - B}_P) \, dV
\end{equation} 
is the helicity of the current-carrying part of the field \citep{1999PPCF...41B.167B} and $H_R$ is the relative helicity as above. We define the 2.5D version of $H_J$ analogously to $H_R^{(2.5D)}$, thus we consider the ratio $\lvert H_J^{(2.5D)}/H_R^{(2.5D)} \rvert$ with
\begin{equation}
    H_J^{(2.5D)} = \int_0^1\int_{-1}^1({\bf A}-{\bf A}_P^{(2.5D)})\cdot({\bf B} - {\bf B}_P^{(2.5D)})\,\mathrm{d}x\mathrm{d}z.
\end{equation}

\subsubsection{Ratios}

We also consider the ratios between the above quantities, generally defined such that the ratios are independent of the magnetic field strength.

\section{Results}  \label{sec:results}

\subsection{Qualitative Behaviour}

We first illustrate the two types of eruption that can occur in the system.

\subsubsection{Zero Photospheric Diffusion}

When we have no photospheric diffusion ($\eta_0 = 0$), the effect of the shearing causes the arcade footpoints to move only in the $y$ direction. The magnetic energy increases as the system is no longer in a relaxed state, but it is not possible for flux ropes to form in the low corona as the footpoints are not brought closer together and there is no flux cancellation. In this case, the only free parameter is the friction coefficient $\nu_0$.

We choose $\nu_0$ of the order unity, in our code units. In general, the equivalent friction coefficents used in 3D global magneto-frictional simulations \citep[e.g.,][]{2016A&A...594A..98Y} are slightly smaller than this if directly compared, but this is by no means a precise measurement. There is thus a compromise between these more realistic values and the significant improvements in computational speed gained from increasing $\nu_0$. Altering the domain shape also has a significant effect on the ideal $\nu_0$, but for our $(x,z) = (2:1)$ ratio setting $\nu_0 \approx 0.5$ produces realistic behaviour. For $\nu_0$ significantly larger than this, the frictional relaxation is unrealistically slow relative to the footpoint shearing.

\begin{figure}[ht]
\includegraphics[scale=0.97]{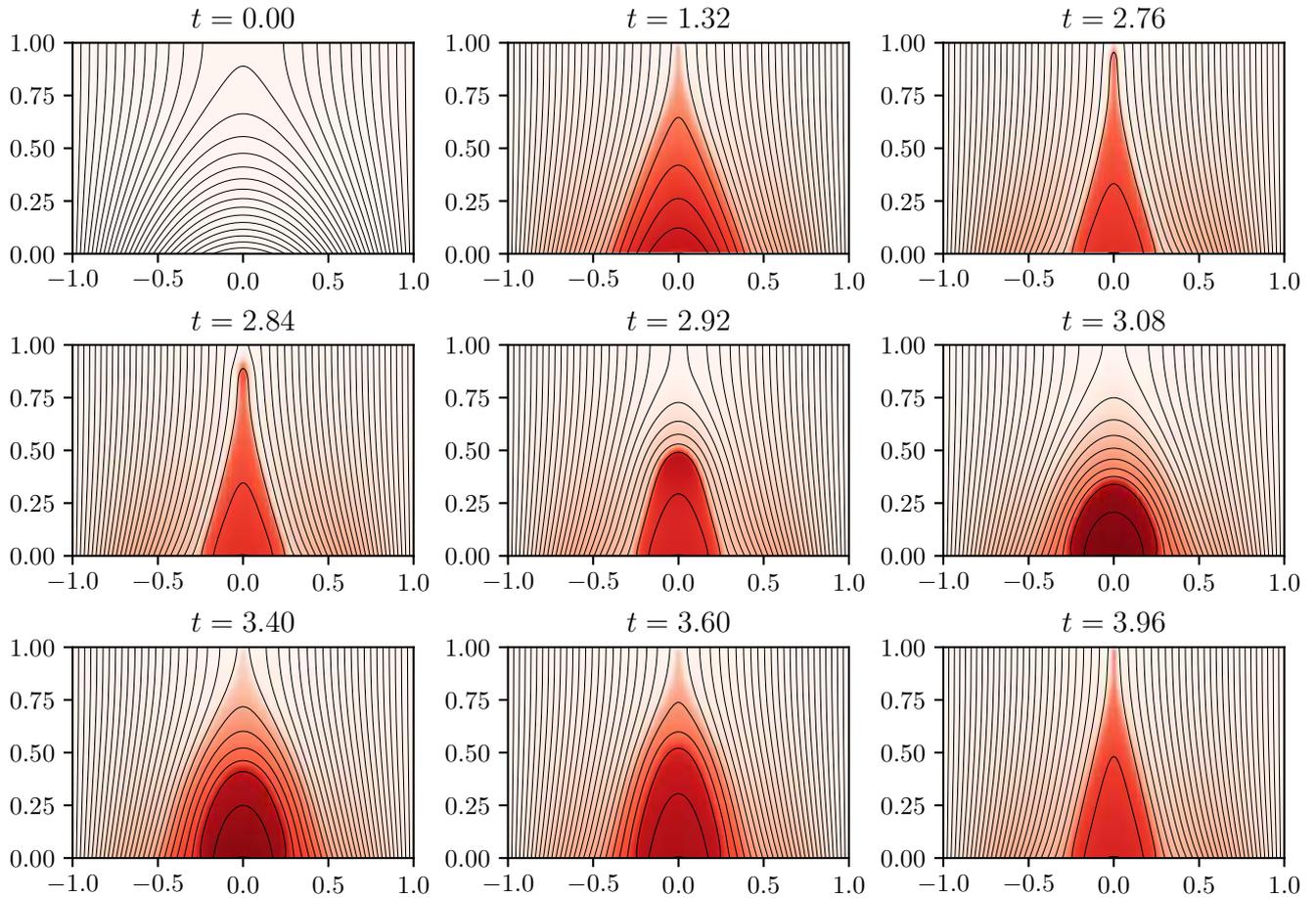}
\caption{Snapshots showing the shearing of the potential arcade resulting in an arcade eruption and subsequent reformation preceding another eruption at $t=4$. The black lines are magnetic field lines projected onto the $(x,z)$ plane, and the heatmap represents the magnetic field strength into the page (in the $y$ direction). In this simulation, $\eta_0=0$ and $\nu_0 = 0.5$.}
\label{fig:type2}
\end{figure}

Consequently, our parameter study is confined to $0.5 < \nu_0 < 2$. For these $\nu_0$ values, the magnetic arcade becomes more sheared in the $y$ direction, as expected. After a certain period during which the magnetic energy and open flux increase gradually, we observe fast magnetic reconnection at the top of the domain accompanied by a sharp decrease in these diagnostics. This is what we refer to as an ``arcade eruption'' \citep{1995ApJ...438L..45L}. A sequence of snapshots of one of these eruptions is shown in Figure \ref{fig:type2}. The first eruption is shown in the centre pane at time $t=2.92$. 

\begin{figure}[ht]
\includegraphics[scale=0.97]{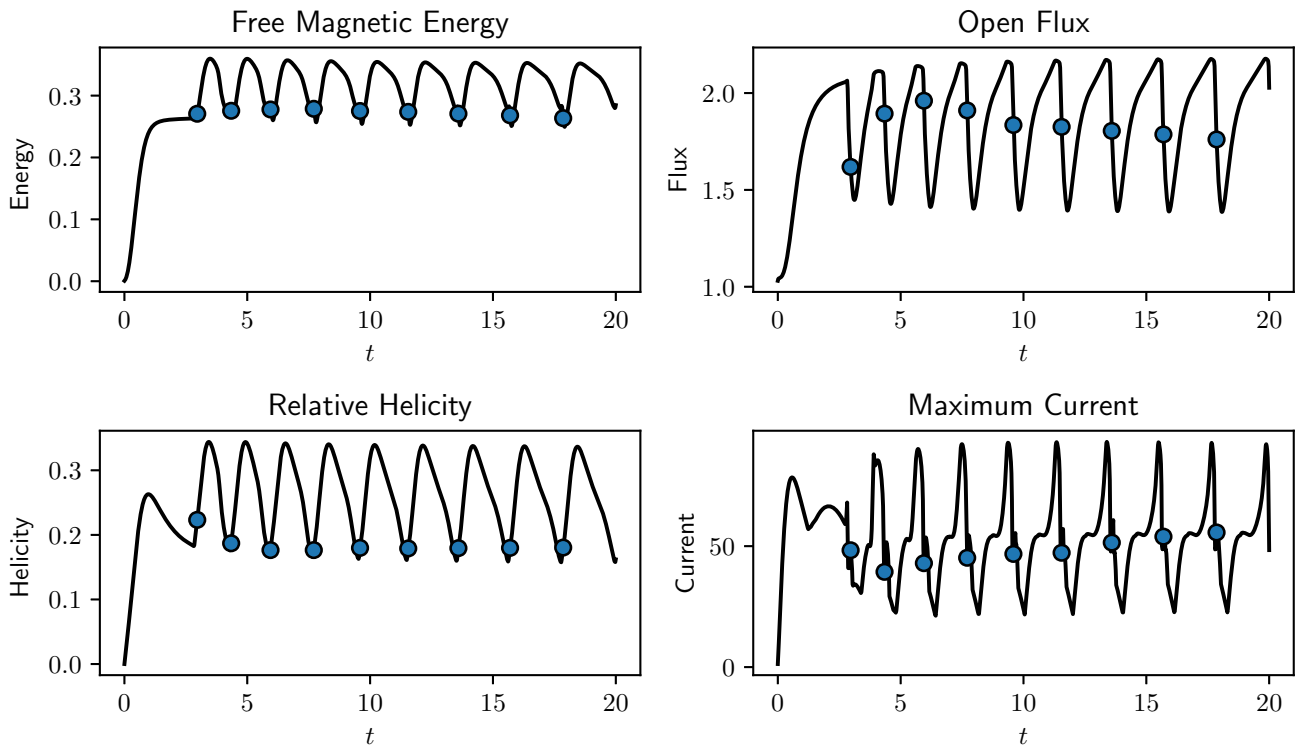}
\caption{A selection of diagnostics for a simulation exhibiting repeated arcade eruptions, which are represented by blue circles. In this simulation, $\eta_0=0$ and $\nu_0=0.5$. The time of an eruption is taken to be the midpoint in time between the maximum and minimum open flux values either side of the eruption.}
\label{fig:type2diags}
\end{figure}

In general this process then repeats, leading to periodic eruptions occuring every time unit or so, as shown clearly by the oscillation of the diagnostic measures in Figure \ref{fig:type2diags}. We observe that the arcade becomes more sheared up to time $t=2.76$, at which point there is an eruption, denoted by a blue circle. The process then repeats, building up to another eruption at $t=4$ and so on, with the times of eruptions (as determined by the mid-point of the  drop in open flux) shown by blue circles.

During an arcade eruption there is a rapid decrease in open flux as the reconnection at the top of the domain results in the closing down of field lines at the top of the arcade. The corresponding decrease in magnetic energy occurs as these newly closed field lines are in a more potential state immediately after the eruption, although the free energy has a non-zero ``floor'' because the background field is non-potential due to the outflow velocity. The peaks in current during an eruption occur at the top boundary at the current sheet that temporarily forms above the PIL. The relative helicity does not follow the same pattern as the other diagnostics and instead increases during an eruption.
This increase results from the fact that the newly-closed potential field lines at the top of the arcade now have a mutual ``linkage'' with the still sheared field lines beneath.

\subsubsection{Nonzero Photospheric Diffusion}

When the photospheric diffusion $\eta_0$ is nonzero, we observe the formation of flux ropes. These are characterised by twisted bundles of magnetic field lines that do not meet the boundary of the domain. As a result, the flux ropes are effectively infinitely long as there is no variation in the $y$ direction. At the same time, we observe similar periodic buildups and releases in magnetic energy to the case with no photospheric diffusion. These overlying arcade eruptions do not affect the amount of poloidal magnetic flux within the flux rope, but they do cause the rope to oscillate vertically.

However, in addition to these eruptions, we also observe more significant events whereby the flux rope itself erupts, which in a 2D setting causes it to move rapidly upward out of the domain. Unlike with arcade eruptions, we observe magnetic reconnection below the flux rope, between it and the polarity inversion line on the lower boundary. There is a significant decrease in free magnetic energy, open flux and relative helicity, and in most cases there is no longer any flux contained within the rope after an eruption (although this is not always the case).

\begin{figure}[ht]
\includegraphics[scale=0.97]{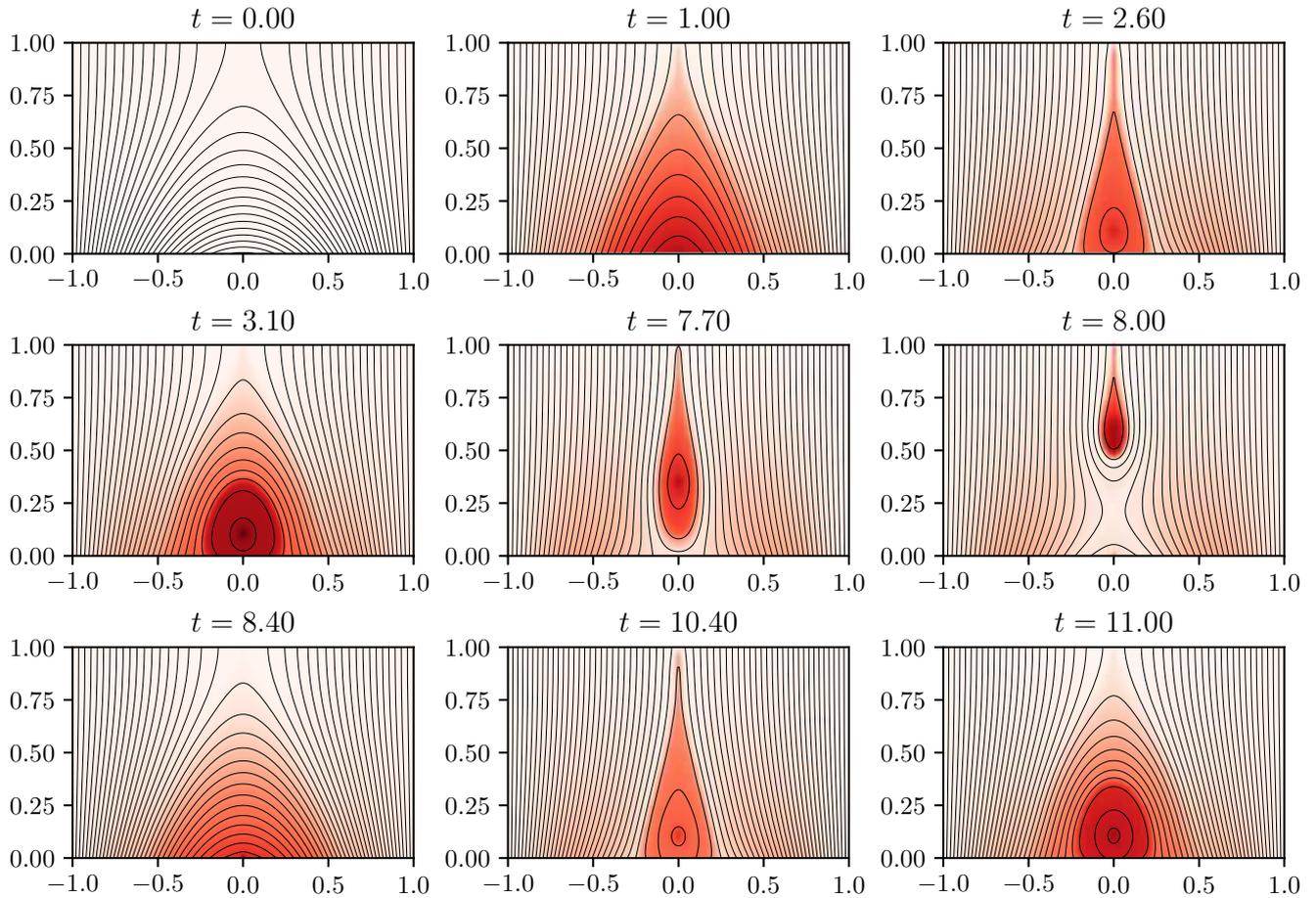}
\caption{Snapshots showing the formation of a flux rope and its subsequent eruption at time $t=8$. This is followed by the formation of a second flux rope, which experiences an arcade eruption at $t=15.8$, before the process repeats. The black lines are magnetic field lines projected onto the $(x,z)$ plane, and the heatmap represents the magnetic field strength into the page (in the $y$ direction). In this simulation,  $\eta_0 = 7 \times 10^{-3}$ and $\nu_0 = 0.6$.}
\label{fig:type1}
\end{figure}

\begin{figure}[ht]
\includegraphics[scale=0.97]{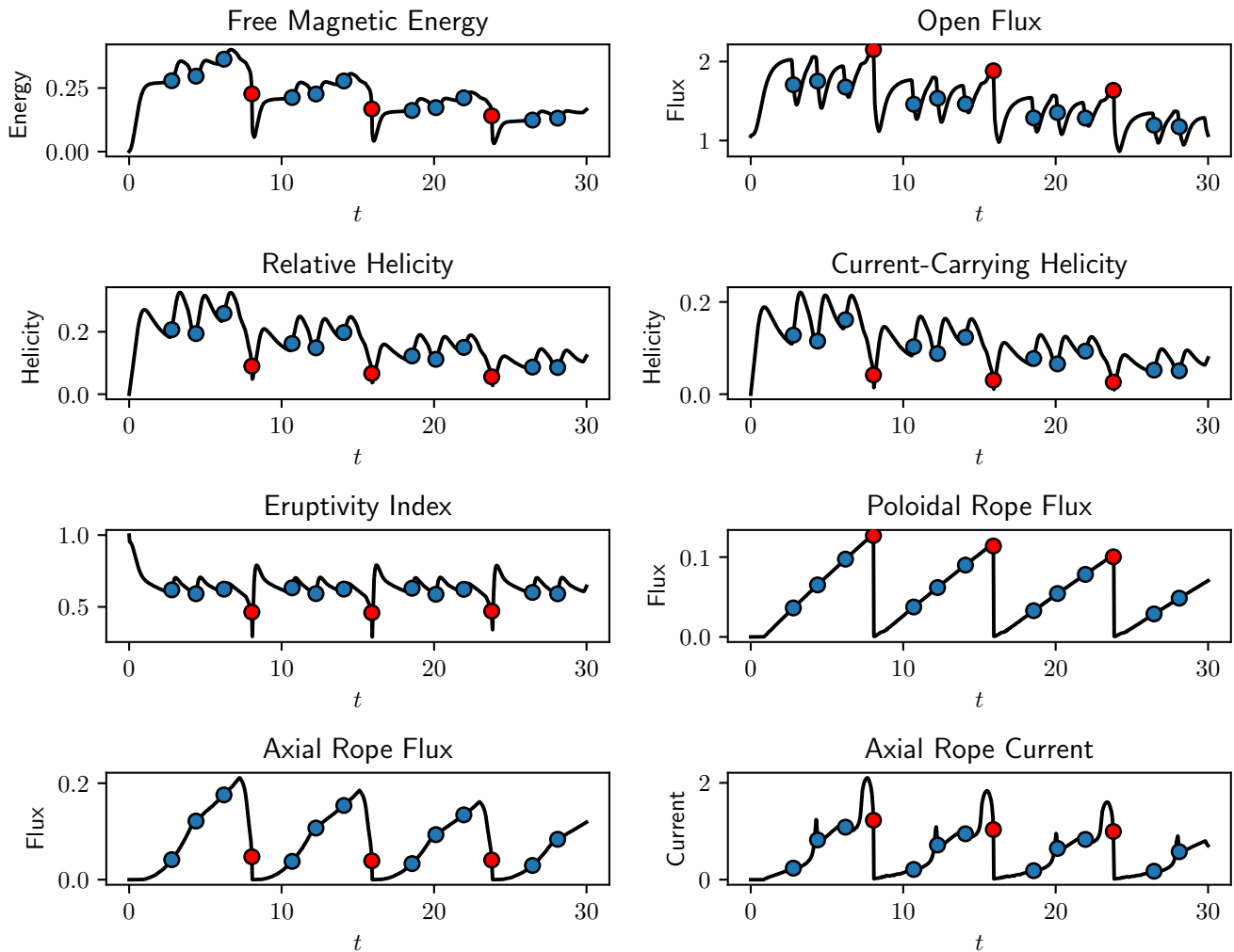}
\caption{A selection of diagnostics for a simulation exhibiting flux rope eruptions (red circles) interspersed with multiple arcade eruptions (blue circles).  In this simulation,  $\eta_0 = 7 \times 10^{-3}$ and $\nu_0 = 0.6$. The time of a flux rope eruption is taken to be the time of the maximum poloidal rope flux before this rapidly decreases.}
\label{fig:type1diags}
\end{figure}

Snapshots showing a flux rope eruption are shown in Figure \ref{fig:type1}, and diagnostics corresponding to the same simulation in Figure \ref{fig:type1diags}. Here the red circles denote the flux rope eruptions, shown at the time when the poloidal rope flux drops. In most cases, after a flux rope eruption a new rope will form and the process will repeat, although depending on the specific parameters the system may just relax into a steady state. Compared to Figure \ref{fig:type2diags}, Figure \ref{fig:type1diags} includes new plots of the rope flux and current as well as the current-carrying helicity $H_J^{(2.5D)}$ and eruptivity index $\lvert H_J^{(2.5D)}/H_R^{(2.5D)} \rvert$. The figure shows both the poloidal/in-plane flux and the axial/out-of-plane flux, contained within the region of infinitely-long magnetic field lines. In our model, prior to eruption, there is no null point beneath the rope. So the poloidal rope flux is simply the difference in the vector potential, $\max(A_y(x,z)) - A_y(0,0)$.

The effect of a flux rope eruption on the diagnostic measurements is more significant than an arcade eruption. We still observe rapid decreases in open flux and magnetic energy, but unlike with arcade eruptions there is now a significant decrease in the relative helicity during the eruption, rather than an increase. Arcade eruptions do not affect the poloidal flux in the rope at all as can be seen in Figure \ref{fig:type1diags}, but naturally as the rope no longer exists after a flux rope eruption these fluxes -- as well as the axial rope current -- immediately fall to zero, before increasing again if flux cancellation is ongoing. The axial rope current and axial rope flux are affected by arcade eruptions, but not significantly. The eruptivity index will be discussed below.

\subsection{Dependence on Photospheric Diffusion}

In the parameter study we primarily focus on the effect of changing the photospheric diffusion parameter $\eta_0$. For very small $\eta_0$ the system exhibits periodic arcade eruptions as described above, and the flux ropes form too slowly to erupt within the timeframe of the simulation, which we set at $50$ time units (corresponding to several years on the real Sun).
It is likely, however, that these ropes will eventually erupt. 

When the photospheric diffusion is in the range $2 \times 10^{-3} < \eta_0 < 1 \times 10^{-1}$ we observe both arcade and flux rope eruptions before $t=50$. For runs where $\eta_0$ is not too large, there are several arcade eruptions between each flux rope eruption that cause the rope to oscillate. At some point, seemingly arbitrarily, one of these arcade eruptions coincides with reconnection below the rope and causes the flux rope to move rapidly upwards and out of the top of the domain. Usually after one of these eruptions the system returns to a potential-like state and the process repeats, although the out-of plane magnetic strength usually diminishes after each eruption. This repeats until the system relaxes to a steady state, with no rope present.

\begin{figure}[ht]
\includegraphics[scale=0.97]{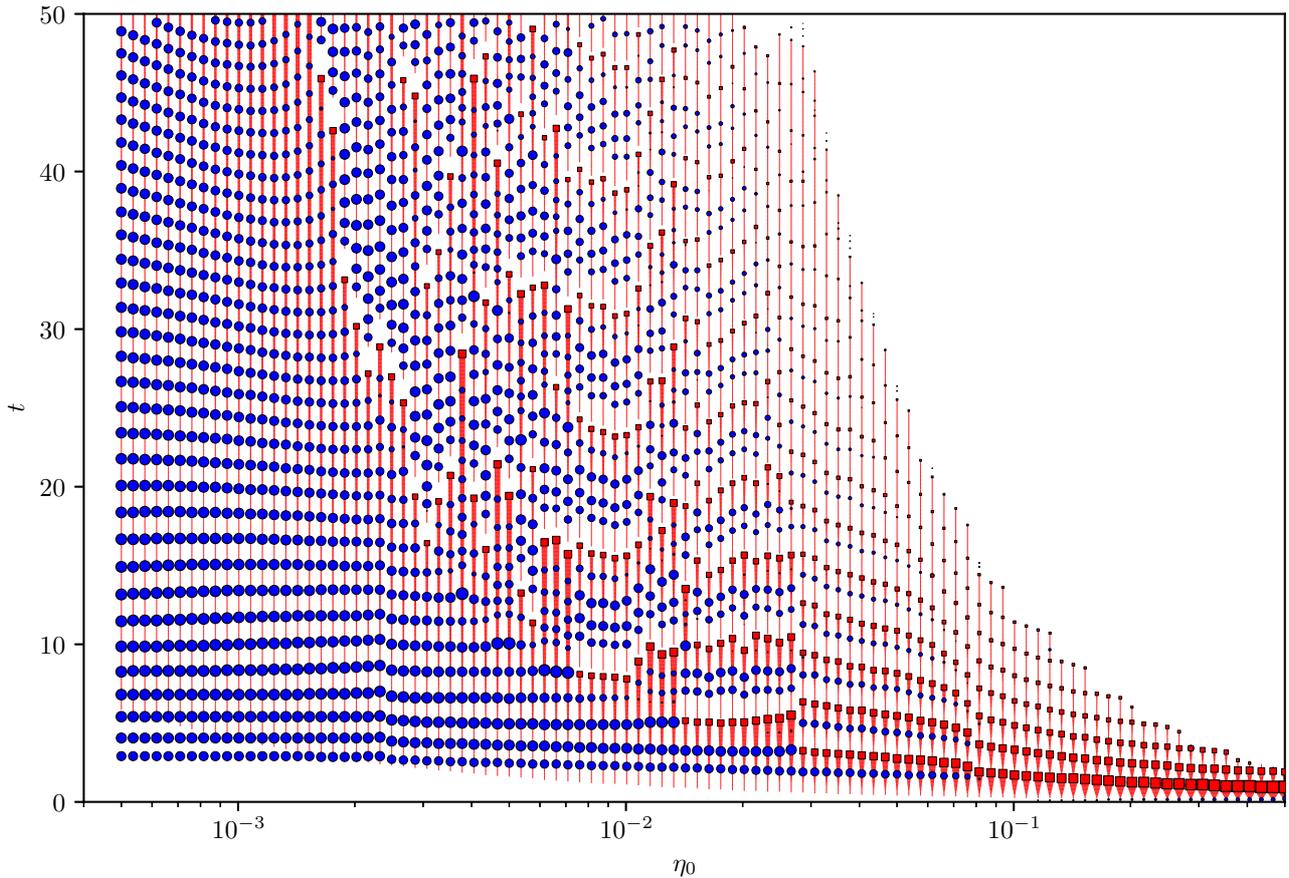}
\caption{Overview of 100 flux rope simulations, for $\nu_0 = 0.5$ and varying $\eta_0$. Each simulation (for a given $\eta_0$) is represented by a vertical red line, and the thickness of this line is proportional to the poloidal flux in the rope at that time. Arcade eruptions are represented by blue circles and flux rope eruptions are represented by red squares. The size of these points is proportional to the decrease in open flux and poloidal rope flux during the eruption, respectively.}
\label{fig:parastudy}
\end{figure}

Varying the value of the photospheric diffusion ($\eta_0$) produces a wide range of flux rope behaviours. Figure \ref{fig:parastudy} shows 100 simulations with $\nu_0 = 0.5$ and varying $\eta_0$. For low $\eta_0$ (the left side of the figure), we observe regular arcade eruptions (indicated by the blue circles) similarly to the cases with no flux ropes. Small ropes do form in this region (indicated by the red lines), and it is likely that they would erupt completely given a sufficiently long integration time.

In the region $\eta_0 > 2 \times 10^{-3}$ we observe flux rope eruptions before $t=50$ (the end of the simulations). In general, these eruptions occur sooner as $\eta_0$ increases, although there is not a simple, continuous dependence of the eruption time on $\eta_0$ for all $\eta_0$. They are usually preceded by a number of arcade eruptions, varying in magnitude. In most cases, after a flux rope eruption a second flux rope forms, and we observe further eruptions of both types.

We observe that the time of the first flux rope eruption, if any, is negatively correlated to the diffusion $\eta_0$. This is visible in the pattern of red squares in Figure \ref{fig:parastudy}, which roughly follow a curve from the top-left of the diagram, flattening out as $\eta_0$ increases. We are unsure whether this curve tends to an asymptote at $\eta_0 = 0$ or at a finite, minimum value of $\eta_0$ below which flux rope eruptions do not occur. The time period between successive flux rope eruptions is also roughly negatively correlated to $\eta_0$.

For $\eta_0 \gtrapprox 0.1$ the flux ropes form and erupt very quickly, and there is no time for arcade eruptions (which have a frequency roughly independent of $\eta_0$). The flux ropes are small but have a high poloidal flux (resulting in the thicker vertical lines on Figure \ref{fig:parastudy}).

\subsection{Diagnostic Measurements as Predictors of Eruptivity}
\label{sec:predicts}
Here we aim to quantify which, if any, of the considered diagnostic measurements in Section \ref{sec:diags} can be used as a proxy for the eruptivity of a flux rope. In this section we only consider flux rope eruptions and not arcade eruptions, which are present at almost all times. We calculate the usefulness of each of the diagnostic measures using a probabilistic approach.

The data used here are from 500 simulation runs up to time $t=50$, covering the parameter space $5 \times 10^{-4} < \eta_0 < 0.5$ and $0.5 < \nu_0 < 2.0$. For reference, a realistic value for $\nu_0$ is likely of order unity. For $\nu_0 > 2$, the relaxation is unrealistically slow relative to the driving, and eruptions become very frequent and less well-defined, with flux ropes not having time to form properly. Each simulation logs $5000$ data points, for a total of $2.5 \times 10^6$ measurements for each diagnostic. We separate each point according to whether it precedes a flux rope eruption, within a certain specified time cutoff.

With multiple flux rope eruptions observed within each simulation, we are able to observe a large distribution of eruption magnitudes and rope sizes/strengths. None of the diagnostics alone are enough to reliably predict imminent eruptions. The magnetic energy, helicity, and open flux all decrease significantly during flux rope eruptions, but in the preceding time interval there is no significant indication of an imminent eruption, and in fact most of the variation in these quantities is due to arcade eruptions. In contrast, the rope fluxes and currents build up from zero with each new rope, until they reach a threshold at which there is an eruption, making them more promising predictors. However, this threshold does not have a constant value, as we can see by observing in Figure \ref{fig:parastudy} that there is large variation in the size of flux rope eruptions. Figure \ref{fig:type1diags} also shows that the peak levels of the rope fluxes and currents tend to reduce for each successive eruption, even in a single simulation.

\begin{figure}[ht]
\includegraphics[scale=0.97]{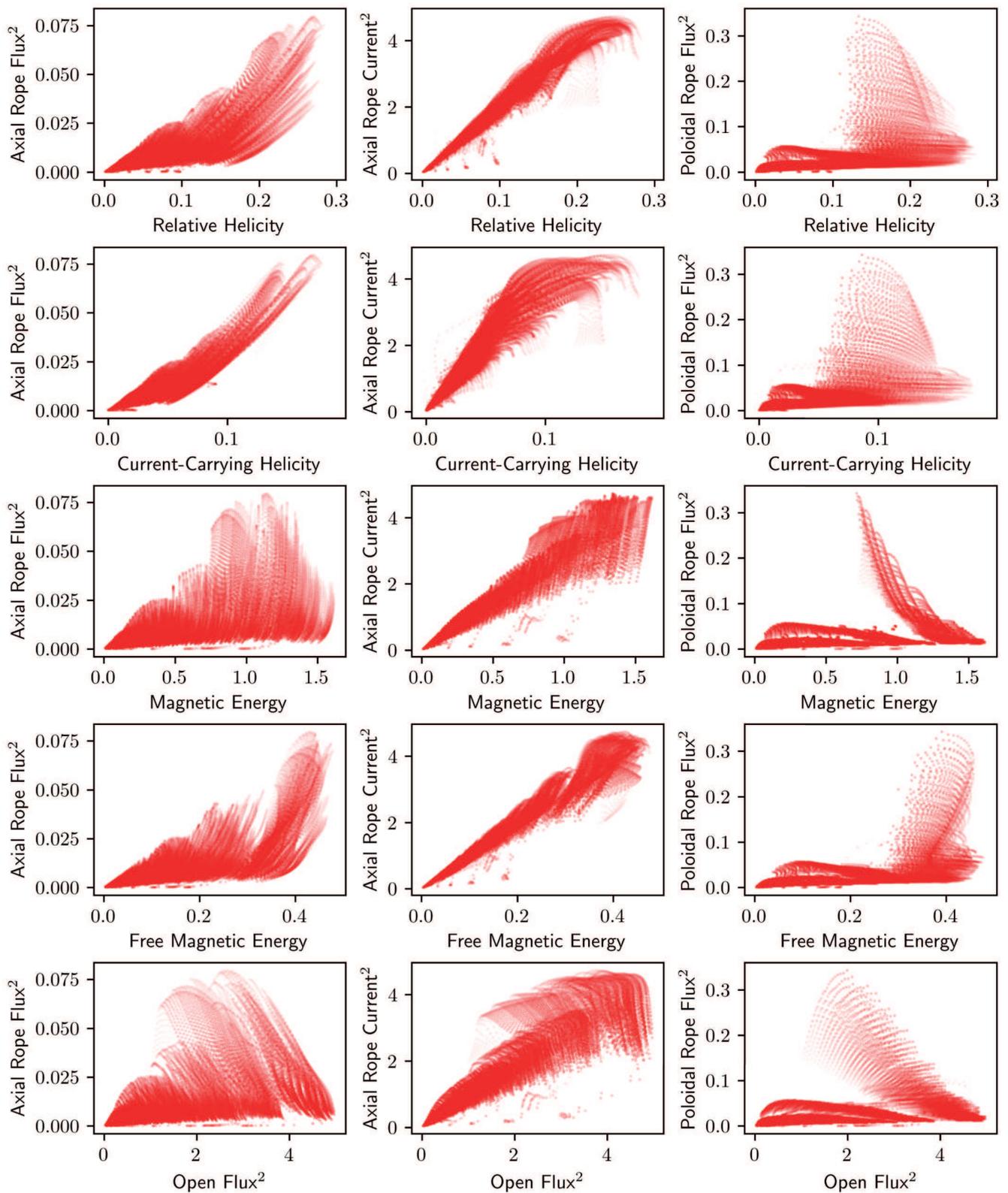}
\caption{Pairwise scatter plots of the diagnostic values for all snapshots with later flux rope eruptions, in order to establish whether the ratios of the diagnostics are good predictors. The sizes of the points are weighted based on proximity to the eruption, such that the larger points are close to eruptions and vice versa.}
\label{fig:ratios}
\end{figure}

However, by calculating the ratios of (squared) rope fluxes and current to the other diagnostics, we observe that some of these ratios do indeed appear to have an eruptivity threshold with a roughly constant value. To motivate this observation, Figure \ref{fig:ratios} shows scatter plots of the points immediately preceding eruptions, with the appropriate diagnostics plotted against one other. The diagnostics on each axis are weighted to be proportional to the square of the magnetic field strength.

In particular, we observe good linear correlations between the square of the axial rope current immediately prior to eruption, and both the magnetic energy and relative helicity of the system. This suggests that the ratios ${I_{y}^2}/{H_R}$, ${I_{y}^2}/{H_J}$, ${I_{y}^2}/{E_M}$ and ${I_{y}^2}/{E_F}$, where $I_y$ is the axial rope current and $E_M, E_F$ are the total and free magnetic energies respectively, should be good indicators of an imminent flux rope eruption. These particular ratios seem to work because the denominators are relatively steady between each successive flux rope eruption, but have a different level each time (this is visible in Figure \ref{fig:type1diags}). Since $I_y^2$ increases prior to a flux rope eruption, the ratios also increase but reach a common, normalized threshold between different eruptions. Note that $H_R$, $H_J$, $E_M$ and $E_F$ all include important contributions from the sheared background field surrounding the rope. Quantities derived only from the rope itself are not found to be successful denominators.

\begin{figure}[ht]
\includegraphics[scale=0.97]{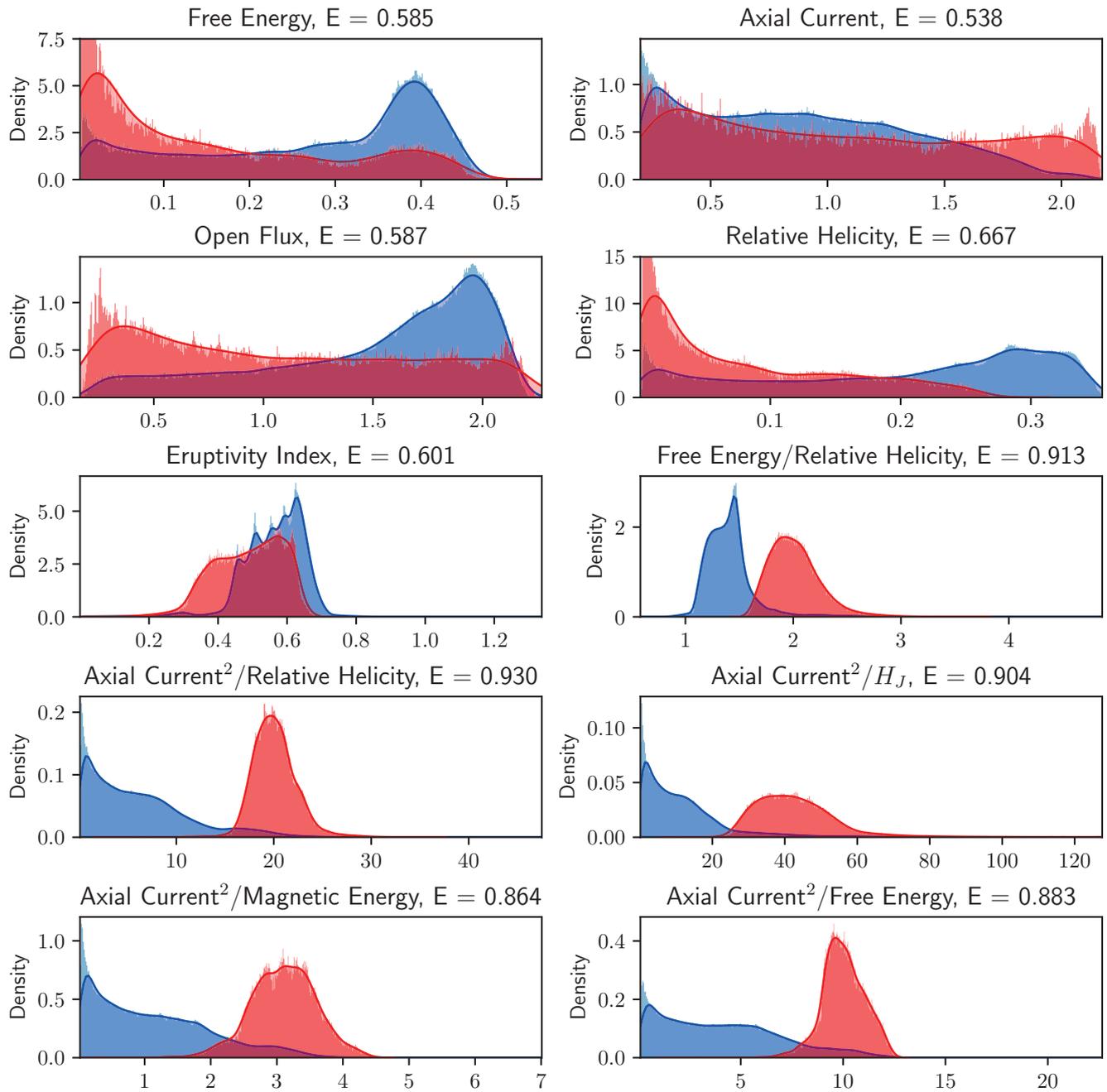}
\caption{Histograms of five of the diagnostic parameters, and five of the ratios between them. The points that precede an eruption within $t=0.3$ time units are shown in red, and those that do not are shown in blue. The diagnostics with less red/blue overlap are better predictors of eruptions and vice versa. The eruptivity skill score $E$ is  given for each diagnostic. The curves are normalised to have an area of unity.}
\label{fig:histograms}
\end{figure}

The distributions of eruptive and non-eruptive points for these ratios and a selection of the diagnostics are shown in Figure \ref{fig:histograms}. At any point in the simulations when a flux rope is present, we check whether the rope will erupt within a time cutoff $t=0.3$ (roughly equivalent to $8$ days). If so, the diagnostics at that time are added to the histograms coloured red. The points corresponding to ropes that will not erupt are coloured blue. The areas under the histograms are corrected for an equal weighting of eruptive and non-eruptive ropes, but in principle this could be altered if the overall prior/uninformed probability of eruption were known. Thus the diagnostics with least red/blue overlap are better predictors of eruptivity than those with a large overlap. 

This can be quantified by assigning a ``probability'' of eruption to each value of each diagnostic. This is essentially the height of the red histogram curve on Figure \ref{fig:histograms} divided by the combined height of both curves, and is output as a number $0 < P_e < 1$ for each data point. The probability $P_e$ can then be compared against whether the rope will actually erupt or not to produce a skill score, defined as 
\begin{equation}
    E = \frac{\sum_{\substack{ \rm Eruptive \\ \rm Points}} P_e + \sum_{\substack{\rm Non - \\ \rm Eruptive \\ \rm Points}}  (1 - P_e)}{{\rm Total \, Number \, of \,Points}}.
\end{equation}
If a diagnostic is a perfect indicator of eruptivity then it would have a skill score $E=1$. If the diagnostic is a no better predictor than random chance then it would have skill score $E=0.5$. The values of $E$ for each diagnostic are given in the headings on Figure \ref{fig:histograms}. We observe that the free energy and open flux are not suitable predictors, as they are little better than chance. The relative helicity and eruptivity index are slightly better, correctly predicting around $60\%$ of eruptions within the time frame. We note, however, that these diagnostics are inversely correlated to the probability of eruption. This is likely because eruptions are more frequent for lower values of the helicity, open flux and energy, as smaller ropes tend to erupt more frequently in our model.

In contrast, the ratios of the squared axial current to the helicities and energies are excellent predictors in this time frame, all with skill scores above $E=0.86$ and significantly better than the constituent diagnostics on their own. The correlations are also all positive, indicating that these ratios increase until they reach a given threshold -- represented by the peaks in the red curves on Figure \ref{fig:histograms}. The best predictor for $t=0.3$ before an eruption (the time at which the predictors are best) is ${I_{y}^2}/{H_R}$, with a skill score of $E=0.93$, or a predictive accuracy of $93.0\%$.

If we disregard the effect of arcade eruptions, the free energy and helicity are roughly constant during the periods between successive flux rope eruptions. However, the ratio between these quantities has also been identified as a suitable predictor, as seen in Figure \ref{fig:histograms}. Immediately before an eruption the relative helicity decreases significantly, before the corresponding decrease in free energy, and thus there is a brief peak in the ratio between the two quantities.

\begin{figure}[ht]
\includegraphics[scale=0.97]{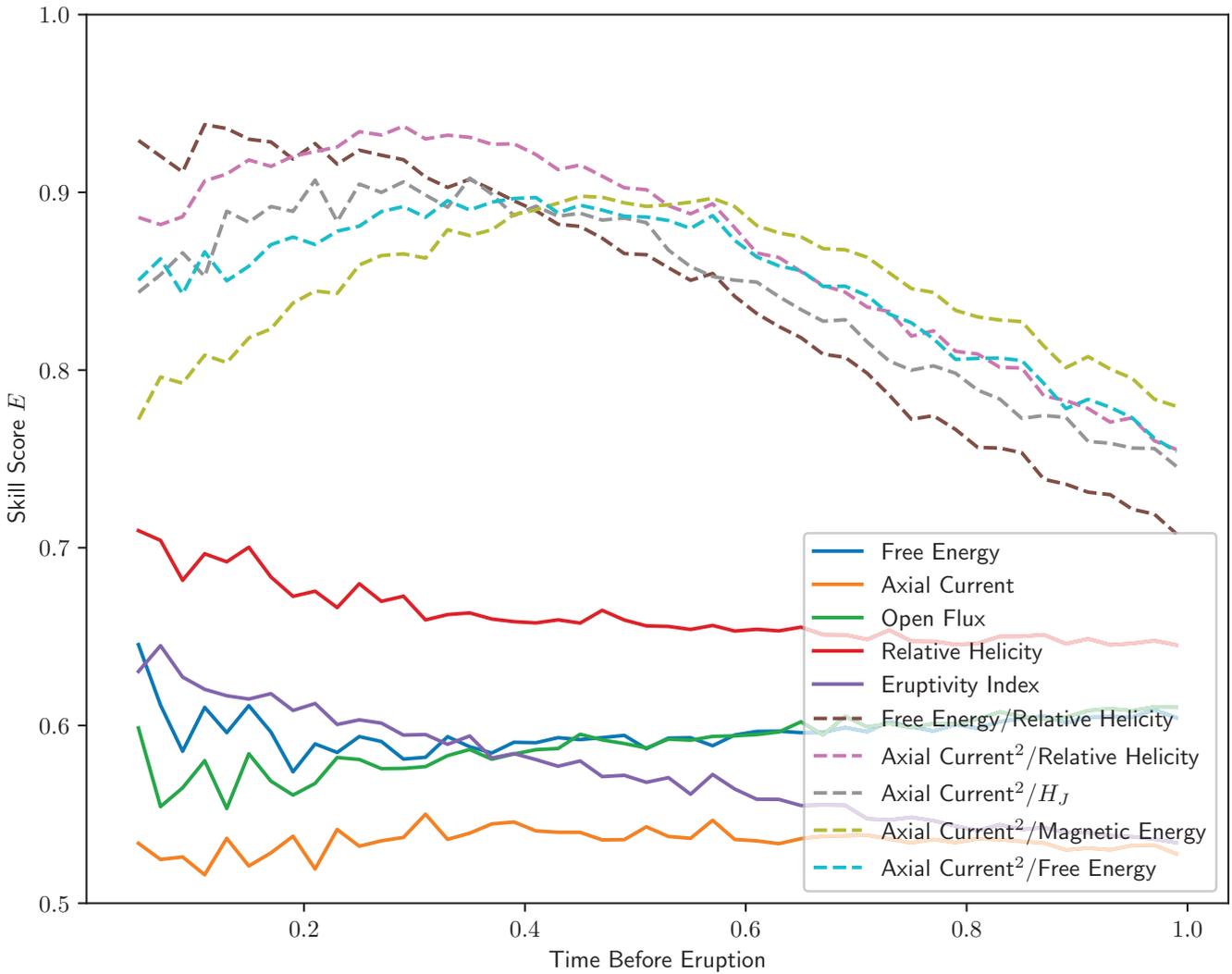}
\caption{Variation of the skill scores for each of the measured quantities, depending on the time cutoff within which an eruption must occur. The raw diagnostics and eruptivity index are plotted as solid lines, and the ratios are plotted as dashed lines.}
\label{fig:eplot}
\end{figure}

The chosen time cutoff naturally has an effect on the skill scores of each of these quantities. This is illustrated in Figure \ref{fig:eplot}, where the skill scores for each of the quantities are plotted against the time cutoff, up to $t=1.0$ before the eruption. Of the raw diagnostics, the relative helicity is consistently the best predictor, with skill scores higher than $E=0.65$ at all times. At almost all times the eruptivity index does not perform as well, and indeed becomes little better than chance for time cutoffs approaching $t=1.0$. The free energy and open flux fare little better, with skill scores around $E=0.6$.

In contrast, the ratios of the axial current to the chosen diagnostics perform consistently well, although their skill score decreases as the time cutoff increases. The quantities $E_{F}/H_R$ and ${I_{y}^2}/{H_R}$ perform best at small time cutoffs, with maximum skill scores of $E=0.938$ and $E=0.937$ respectively. The ratios of the axial current to magnetic energy also perform well, and indeed for time cutoffs greater than $t=0.5$ these ratios perform equally well or better than the other ratios. The maximum skill scores for most of the ratios occurs at a time cutoff of around $t=0.3$, which corresponds to several days in reality. As such these predictors have potential for predicting solar eruptions a useful time before they occur.

\section{Discussion}  \label{sec:discussion}

In an extensive parameter study, we have analysed the behaviour of two-dimensional magnetic flux ropes to examine which, if any, properties of the system can be used to predict whether or not the rope will erupt, an event which in reality is likely to cause a CME.
For our simulations we used the magneto-frictional model rather than full MHD, reducing computational complexity further and allowing us to run thousands of simulations over a wide parameter space. We observe repeated arcade eruptions as in equivalent 3D simulations, as well as the formation of flux ropes. The main focus of our results is the prediction of the eruption of the flux ropes, rather than eruptions within the overlying arcade.

We have observed that the behaviour depends greatly on the values of the magneto-friction coefficient $\nu_0$ and the photospheric diffusion $\eta_0$. In particular, flux ropes form and erupt more quickly for higher $\eta_0$, whereas the frequency of arcade eruptions within the overlying field is roughly independent of $\eta_0$. By comparing the probability of a rope erupting within a certain time to a number of diagnostic measurements, we have assigned a ``skill-score'' to each of them and the ratios between them. Of particular interest are the relative helicity (which we have newly defined for 2.5D systems) and its current-carrying component, $H_J$. The ratio of $H_J$ to the total relative helicity constitutes the eruptivity index of \citet{2017A&A...601A.125P}.

We found that none of the diagnostics considered were by themselves good predictors of eruptivity, with skill scores not significantly greater than random chance. Of note, the eruptivity index has similar predictive skill to the other diagnostics, and is in fact negatively correlated to the likelihood of a flux rope eruption, unlike in the previous MHD simulations \citep{2017A&A...601A.125P, 2018ApJ...863...41Z}. However, when we consider the ratios between the diagnostics, weighted so as to be independent of the overall magnetic field strength, we find that certain of these can be good predictors.

Of particular note are the ratios with the axial current as the numerator, and either a helicity or energy measure as a denominator. The axial current indicates the ``strength'' of the rope, and the helicity/energy is in effect a measure of the strength of the background field. These denominator are roughly constant in between eruptions, whereas the axial current steadily increases as the rope becomes larger. Upon the ratio reaching a certain threshold, the rope will erupt. Notably, this threshold appears to be independent of both $\eta_0$ and $\nu_0$. The ratio of free energy to relative helicity is also a very good predictor for eruptions in the immediate future, as a rapid decrease in relative helicity is often followed by an eruption.

\subsection{Effect of Background Magnetic Field on the Eruptivity Index} 
\label{sec:backfield}

In this section we propose a straightforward explanation for why \citet{2017A&A...601A.125P} observed a high eruptivity index prior to the flux rope eruption, whereas we do not. We propose that this difference is due to the direction of the overlying magnetic field.
We will consider a simple analytical model, which shows that the eruptivity index will naturally be higher when the background/overlying horizontal field direction is opposite to that of the arcade. 

\begin{figure}[ht]
\includegraphics[scale=0.97]{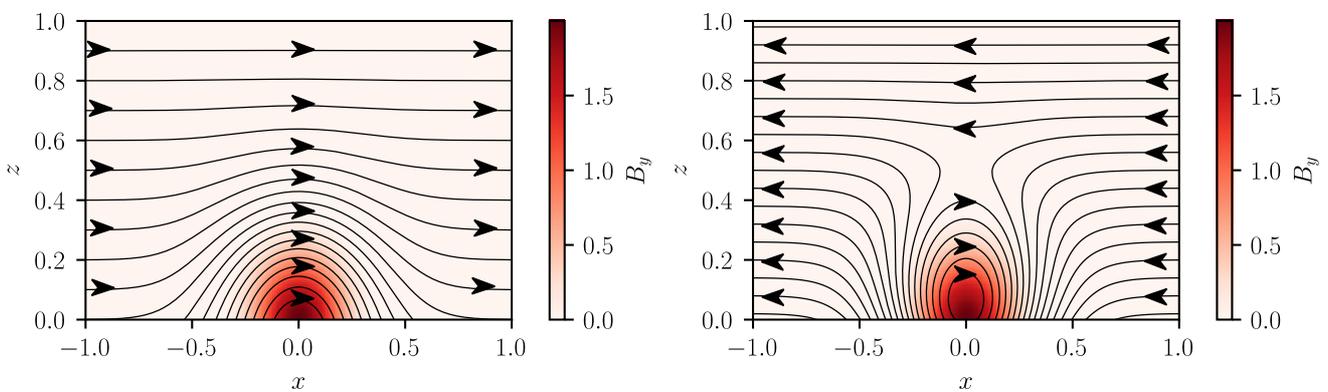}
\caption{Comparison between two magnetic arcades with overlying magnetic fields in opposite directions. The black lines represent the magnetic field projected into the $(x,z)$ plane and the heatmap represents the magnetic field strength out of this plane.}
\label{fig:backfields}
\end{figure}

Two magnetic field configurations are presented in Figure \ref{fig:backfields}. The left pane shows a configuration similar to the fields we generate naturally by shearing a potential field, where the background magnetic field is orientated in the same direction as the arcade. The right pane has the background field in the opposite direction, leading to a magnetic null point above the arcade. When flux ropes emerge into the second type of field, they are more likely to erupt. This was shown clearly by \citet{2017A&A...601A.125P}, who compared simulations with both orientations of the overlying field.
We endeavour to show here that the right-hand field configuration fundamentally results in a higher eruptivity index, for a given sheared field component. 
By contrast, our simulations correspond to the left-hand field configuration, so even though they do erupt, this is not accompanied by a high eruptivity index.

The model magnetic field plotted in Figure \ref{fig:backfields} comes from the analytical expression ${\bf B} = \widetilde{\bf B} + B_0{\bf e}_x$, where
\begin{align}
    \widetilde{B}_x &= 4ze^{-\xi} \label{eqn:bspec1}\\
    \widetilde{B}_y &= 2(1-\xi)e^{-2\xi} \\
    \widetilde{B}_z &= -4xe^{-\xi}, \label{eqn:bspec2}
\end{align}
with $\xi = 4(x^2 + z^2)$. The sheared, out-of-plane component, $B_y$, is fixed, and the only parameter is the strength of the background magnetic field, given by the (constant) parameter $B_0$. The left pane in Figure \ref{fig:backfields} has $B_0 = 0.25$, the right $B_0 = -0.25$. We proceed to observe the dependence on $B_0$ of $H_R^{(2.5D)}$ and $H_J^{(2.5D)}$, as defined in Section \ref{sec:relhel}, and the resultant effect on the eruptivity index $\lvert {H_J^{(2.5D)}}/{H_R^{(2.5D)}}\rvert$.

Following Section \ref{sec:relhel}, in order to calculate the relative helicity we choose a vector potential ${\bf A} = \widetilde{\bf A}(x,z) -B_0z{\bf e}_y$, where $\nabla\times\widetilde{\bf A}=\widetilde{\bf B}$ and $\widetilde{\bf A}$ is independent of $B_0$. Since the $B_0{\bf e}_x$ component of ${\bf B}$ is a potential field, we can similarly choose ${\bf A}_P^{(2.5D)} = \widetilde{\bf A}_P(x,z) -B_0z{\bf e}_y$. It follows that, for these fields, the ``current-carrying'' helicity $H_J^{(2.5D)} = \iint ({\bf A - A}_P^{(2.5D)}) \cdot ({\bf B - B}_P^{(2.5D)}) \, \mathrm{d}x \mathrm{d}z$ has no dependence on the background field $B_0$, because the $B_0$ terms from ${\bf A}$ and ${\bf A}_P^{(2.5D)}$ will cancel. For the relative helicity, however, $H_R^{(2.5D)} = \int_V ({\bf A + A}_P^{(2.5D)}) \cdot ({\bf B - B}_P^{(2.5D)}) \, \mathrm{d}V$, so the two terms add together and it does depend on $B_0$. Thus we can write
\begin{equation}
H_R^{(2.5D)} = H_0^{(2.5D)} - B_0 \iint (2 z {\bf e}_y) \cdot ({\bf B - B}_P^{(2.5D)}) \, \mathrm{d}x \mathrm{d}z, 
\end{equation}
where $H_0^{(2.5D)}$ is the relative helicity with $B_0 = 0$. 
Since $H_0^{(2.5D)}$ and $H_J^{(2.5D)}$ have no dependence on $B_0$, the eruptivity index can simply be expressed as 
\begin{equation}
    \lvert {H_J^{(2.5D)}}/{H_R^{(2.5D)}}\rvert = \left\lvert \frac{H_J^{(2.5D)}}{H_0^{(2.5D)} - B_0 \iint (2 z {\bf e}_y) \cdot ({\bf B - B}_P^{(2.5D)}) \, \mathrm{d}x \mathrm{d}z} \right\rvert.
    \label{eqn:hrat}
\end{equation}
For a sheared field with $B_y$ non-zero (and non-uniform), we can clearly see that there will be a particular background field strength $B_0$ where the eruptivity index will become infinite as the denominator vanishes. For the magnetic field specified in Equations \eqref{eqn:bspec1} to \eqref{eqn:bspec2} the constants take the values
\begin{align}
H_J^{(2.5D)} &\approx -0.0109 \\
H_0^{(2.5D)} &\approx 0.0307 \\
\int_V (2 z {\bf e}_y) \cdot ({\bf B - B}_P^{(2.5D)}) \, dV &\approx -0.2026,
\end{align}
which results in a peak in the eruptivity index at $B_0 \approx -0.15$, when in particular the overlying magnetic field is oppositely directed to the magnetic field in the upper part of the arcade (as in the right pane of Figure \ref{fig:backfields}). By contrast, if the overlying magnetic field has the same direction as that in the arcade, so that $B_0>0$ (as in the left pane of Figure \ref{fig:backfields}), then the denominator of \eqref{eqn:hrat} will not become very small so the eruptivity index will not become large.

In all of our simulations -- where the flux rope is formed by shearing of a pre-existing potential arcade -- the background field has the same direction as that of the arcade, whether or not the flux rope erupts.
Generalising from the analytical model with $B_0>0$, this explains why our eruptions are not preceded by a high eruptivity index.

The simulations of \citep{2017A&A...601A.125P}, which were driven by flux emergence, included cases with both directions of background field. The eruptivity index behaved as predicted by the simple model in this section, but in that case only the cases with oppositely-directed field (and high eruptivity index) erupted.
Our work shows that there is a whole class of eruptions that will not have high eruptivity index owing to the fact that they occur despite having the same direction of overlying field.

\subsection{Implications for space weather forecasting} 

The simplified nature of our 2.5D system means that any quantitative predictions will not be valid in 3D or indeed for any differing domain size or shape. However, the qualitative patterns of behaviour that we observe (such as those of the ratios of axial rope current to overall helicity/energy) should be equally valid in all systems, including global coronal models, where flux ropes are formed by footpoint shearing from differential rotation. Moreover, since these flux ropes are formed by gradual shearing over days to weeks, and are located in the magnetically-dominated low corona, we do not expect that the general conclusions would change significantly if we were to move from the magneto-frictional model to full MHD. For example, it has been shown that the linear stability criteria in magneto-frictional and MHD systems are the same \citep{1986ApJ...311..451C}.
Nevertheless, to make quantitative predictions about specific 3D magnetic configurations on the Sun will require further work to understand how the behaviour of the diagnostics depends on the local coronal magnetic stucture.

\section*{Conflict of Interest Statement}

The authors declare that the research was conducted in the absence of any commercial or financial relationships that could be construed as a potential conflict of interest.

\section*{Author Contributions}

OEKR did all of the calculations and worked out the proof in Section \ref{sec:hr}, under the guidance of ARY who conceived the project. Both authors contributed to writing the paper. We thank the two referees for valuable comments that have greatly improved the paper.

\section*{Funding}
OEKR was supported by a UKRI/STFC PhD studentship, and ARY by UKRI/STFC research grant ST/S000321/1.

\section*{Acknowledgments}

\section*{Supplemental Data}

\section*{Data Availability Statement}
The datasets generated in this study are available from the corresponding author on reasonable request.

\bibliographystyle{frontiersinSCNS_ENG_HUMS} 
\bibliography{fluxropebib}


\section*{Figure captions}

\end{document}